\documentclass[runningheads]{llncs}
\usepackage[T1]{fontenc}
\usepackage{graphicx,verbatim}
\usepackage{booktabs}
\usepackage{amssymb}
 \usepackage{amsmath}
\begin{document}
\title{Subject-Specific Low-Field MRI Synthesis via a Neural Operator}

%
\author{Ziqi Gao\inst{1} \and
Nicha Dvornek\inst{1,2} \and
Xiaoran Zhang\inst{1} \and
Gigi Galiana\inst{1,2} \and
Hemant Tagare\inst{1,2} \and
Todd Constable\inst{1,2}}
\authorrunning{Gao et al.}
%
\institute{Departments of Biomedical Engineering, Yale University \and
Department of Radiology \& Biomedical Imaging, Yale University\\
\email{ziqi.gao@yale.edu}\\}

\maketitle              
\begin{abstract}

Low-field (LF) magnetic resonance imaging (MRI) improves accessibility and reduces costs but generally has lower signal-to-noise ratios and degraded contrast compared to high field (HF) MRI, limiting its clinical utility. Simulating LF MRI from HF MRI enables virtually evaluating novel imaging devices and developing LF algorithms. Existing low field simulators rely on noise injections and smoothing, which fails to capture the contrast degradation seen in LF acquisitions. To this end, we introduce an end-to-end LF-MRI synthesis framework that learns HF to LF image degradation directly from a small number of paired HF-LF MRIs. Specifically, we introduce a novel \textbf{H}F to \textbf{L}F coordinate–image decoupled neural \textbf{O}perator (H2LO) to model the underlying degradation process, and tailor it to capture high-frequency noise textures and dedicated image structure. Experimental results in T1w and T2w MRI demonstrate that H2LO produces more faithful simulated low-field images than existing parameterized noise synthesis model and popular image-to-image translation models. Furthermore, it improves performance in downstream image enhancement tasks, showcasing its potential to enhance LF MRI diagnostic capabilities.


\end{abstract}
\section{Introduction}

Low-field (LF) magnetic resonance imaging (MRI) offers improved accessibility and reduced cost compared to conventional high-field (HF) systems~\cite{campbell2019opportunities,mazurek2021portable,sabir2023feasibility,samardzija2024low,yuen2022portable}, yet generally produces images with lower signal-to-noise ratio (SNR), reduced spatial resolution, and diminished tissue contrast~\cite{arnold2022simulated,figini2020image,javadi2025silico,lin2024zero,man2023deep,o2022vivo}, limiting its clinical utility. Establishing the clinical value of novel LF-MRI devices relies on prospective clinical studies~\cite{campbell2019opportunities,mazurek2021portable,sabir2023feasibility,yuen2022portable}, which typically involve cohorts ranging from tens to over one hundred subjects. Assembling such validation cohorts often takes years~\cite{campbell2019opportunities,mazurek2021portable}, creating a bottleneck for rapid device development and evaluation.

In-silico retrospective analysis~\cite{arnold2022simulated} offers a complementary strategy to accelerate prospective clinical trials and device development by synthesizing subject-specific LF images from publicly available HF datasets. The HF-to-LF transformation can be learned from a small number of paired HF–LF scans, which are considerably easier to acquire than the large cohorts required for prospective validation. Arnold et al. proposed a parameterized noise-and-smoothing model that minimizes discrepancies in global histogram features using three subject pairs \cite{arnold2022simulated}. However, this approach assumes a monotonic intensity transformation and therefore cannot faithfully capture the non-linear, tissue-specific contrast shifts induced by field strength changes~\cite{o2022vivo}. In particular, differential relaxation effects may alter the relative positions of tissue subpeaks (e.g., white matter and gray matter) within the intensity distribution, leading to inaccurate anatomical contrast in the synthesized LF images. Consequently, it has limited capacity to represent broader contrast variations across field strengths (Fig.~\ref{fig:fig2}(a), second column; (d), purple line).


To capture both SNR degradation and contrast changes, we formulate the HF-to-LF transformation as a field-strength–dependent transformation of the underlying MR signal and learn it in an end-to-end manner. Although this problem can be viewed from an image-to-image (I2I) translation perspective~\cite{pix2pix2017,xia2024diffi2i}, HF-to-LF translation differs from conventional pixel-level appearance mappings: it requires modeling how field strength alters the continuous spatial signal, including its frequency content and contrast behavior. In our experiments, popular pixel-wise I2I models~\cite{pix2pix2017,xia2024diffi2i} are unable to simultaneously capture LF textures and accurately model contrast changes (3${rd}$ and 4${th}$ columns of Figure~\ref{fig:fig2}(a)). Therefore, we propose a novel high-frequency preserving 3D neural operator (H2LO), a lightweight coordinate-based network designed to model this transformation while preserving high-frequency structural information, and demonstrate its performance on both T1w and T2w MRI.

Besides LF MRI synthesis, we further demonstrate that H2LO supports LF algorithm development as a faithful data augmentation technique. We consider the LF image enhancement task~\cite{man2023deep,javadi2025silico,lin2024zero,cui2023meta,kim20233d}, which aims to transform LF MRI into HF-like MRI to improve diagnostic utility.
We augment an existing supervised LF image enhancement model~\cite{man2023deep}, by simulating LF MRI from another public conventional MRI dataset, thereby creating additional LF–HF pairs.


In summary, our contributions are threefold. First, to support virtual clinical trials and algorithm development for novel MRI devices, we introduce the first end-to-end framework for LF MRI synthesis. Second, we formulate HF-to-LF translation as an operator learning problem. Building upon a theoretically bounded neural operator architecture \cite{lu2021learning}, H2LO is tailored for preserving high-frequency in 3D volumetric data. Third, we compare H2LO with popular I2I translation networks and an LF simulator in terms of both synthesis quality and downstream utility, demonstrating superior performance on T1w and T2w MRI.

\section{Method}
\vspace{-2pt}
\begin{figure}[t]
    \centering
    \includegraphics[width=0.95\linewidth]{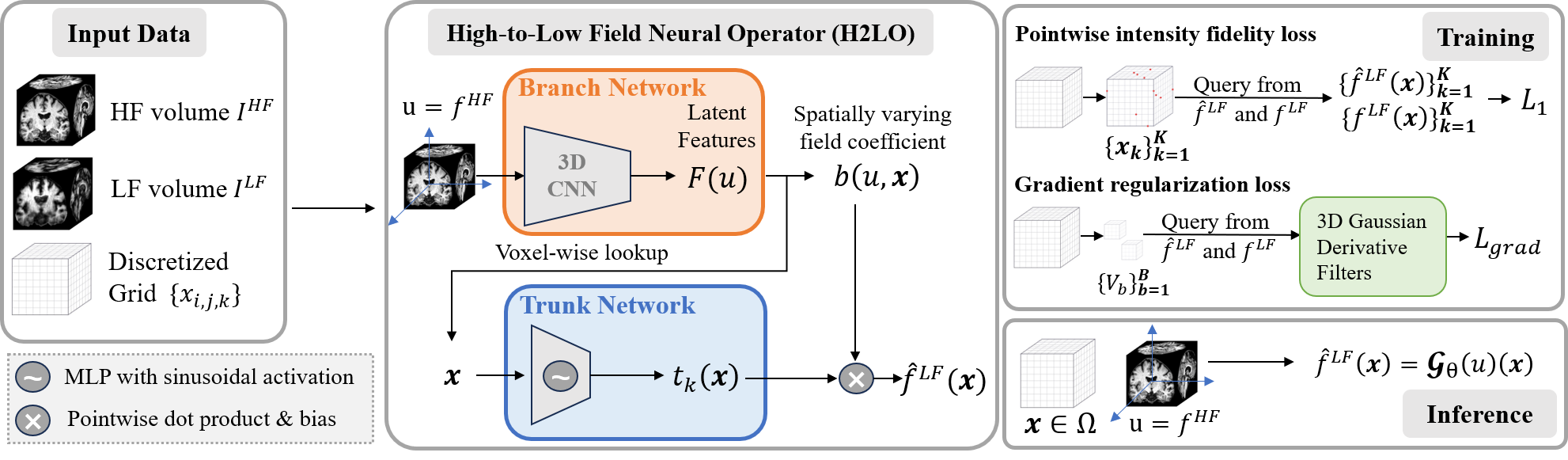}
    \vspace{-5pt}
    \caption{Overview of the High-to-Low field Operator (H2LO) framework. The architecture maps HF MRI volumes to a continuous LF representation using a branch-trunk network optimized via pointwise intensity fidelity and gradient regularization.}
    \label{fig:method_overview}
    \vspace{-10pt}
\end{figure}
\subsection{Problem Formulation}
Let $I^{HF} \in \mathbb{R}^{H \times W \times D}$ and 
$I^{LF} \in \mathbb{R}^{H \times W \times D}$ denote a spatially aligned pair of 
HF and LF 3D MRI volumes acquired at different magnetic field strengths. 
From the perspective of MRI signal formation, image intensity depends on 
field-strength–dependent relaxation properties (e.g., T1 and T2) and 
SNR characteristics. Changing the magnetic field strength therefore 
alters the underlying signal response and its spatial frequency behavior, 
rather than merely modifying voxel-wise intensities. 

We interpret each discretized MRI volume acquired at field strength $E$ as samples from an underlying continuous spatial signal 
$f^{E} : \Omega \subset \mathbb{R}^3 \rightarrow \mathbb{R}$. 
Specifically,
$ I^{HF}[i,j,k] = f^{HF}(\mathbf{x}_{i,j,k}),     
I^{LF}[i,j,k] = f^{LF}(\mathbf{x}_{i,j,k}), $
where $\{\mathbf{x}_{i,j,k}\}$ denotes the sampling grid.

Under this function-space formulation, HF-to-LF (H2L) translation amounts to learning a field-strength–dependent operator
$ \mathcal{T}_{HF \rightarrow LF} : f^{HF} \mapsto f^{LF}. $
Let $u := f^{HF}$ for notation simplicity. We approximate $\mathcal{T}_{HF \rightarrow LF}$ with a neural operator 
$\mathcal{G}_{\theta}$ such that
$
\hat{f}^{LF}(\mathbf{x}) = 
\mathcal{G}_{\theta}(u)(\mathbf{x}), \mathbf{x} \in \Omega.
$

This operator perspective models H2L translation as a mapping between function spaces rather than a fixed-grid voxel regression, aligning the formulation with the continuous nature of MR signal formation.

\subsection{H2LO: High-to-Low Field Operator}
We parameterize $\mathcal{G}_{\theta}$ utilizing branch-trunk factorization following DeepONet~\cite{lu2021learning}. To let local tissue contrast modulate the operator response at each voxel, we extend the original formulation: while DeepONet's branch coefficients are global summaries of the input function, our branch produces spatially varying coefficients by reading from a dense CNN feature field. The resulting operator takes the form
\begin{equation}
\mathcal{G}_{\theta}(u)(\mathbf{x}) = \sum_{k=1}^{P} b_k(u,\mathbf{x})\, t_k(\mathbf{x}) + \beta.
\end{equation}
where $b_k(u,\mathbf{x})$ are spatially varying field coefficients, $t_k(\mathbf{x})$ are coordinate-dependent basis functions, and $\beta$ is a learnable bias term. As shown in Fig.~\ref{fig:method_overview}, our framework comprises (1) a branch network and (2) a trunk network.

\paragraph{Branch network.}
The branch is a 3D image encoder that maps the input volume to a dense feature field $\mathbf{F}(u) \in \mathbb{R}^{P \times H \times W \times D}$. We adopt a lightweight 3D convolutional architecture from SRResNet~\cite{ledig2017photo}. For each query location $\mathbf{x}_{i,j,k}$, the coefficient vector $\mathbf{b}(u,\mathbf{x}_{i,j,k}) \in \mathbb{R}^{P}$ is obtained by reading the feature vector at the corresponding voxel. This produces spatially varying operator coefficients while requiring only a single encoder forward pass per volume.

\paragraph{Trunk network.}
The coordinate basis functions $t_k(\mathbf{x})$ are parameterized by a sinusoidal representation network (SIREN)~\cite{sitzmann2020implicit} with four layers: one input sine layer, two hidden sine layers, and a linear output layer producing $P=128$ outputs. Each sine layer computes $\sin(\omega_0 \mathbf{W}\mathbf{x} + \mathbf{b})$ with SIREN initialization. Sinusoidal activations enable accurate modeling of high-frequency spatial structures, preserving fine anatomical boundaries in the synthesized image.
\vspace{-5pt}

\subsection{Training and Inference}

The total training loss combines an intensity fidelity term and a local gradient regularization term: $ \mathcal{L} = \mathcal{L}_{1} + \lambda_{grad} \mathcal{L}_{grad}. $
The $\mathcal{L}_{1}$ term enforces zero-order fidelity over $N$ randomly sampled voxel coordinates $\{\mathbf{x}_n\}_{n=1}^{N}$: 

$$\mathcal{L}_{1} = \frac{1}{N} \sum_{n=1}^{N} \left| \mathcal{G}_{\theta}(u)(\mathbf{x}_n) - f^{LF}(\mathbf{x}_n) \right|.$$

Reconstructing high-fidelity details requires further regularization of the local behavior of the learned operator. We therefore introduce a first-order consistency term by penalizing discrepancies between spatial derivatives of the predicted and target functions over randomly cropped sub-volumes:
\begin{equation}
\mathcal{L}_{grad} = \frac{1}{B} \sum_{b=1}^{B} \sum_{d \in \{x,y,z\}} 
\left\| 
\partial_d \mathcal{G}_{\theta}(u) - \partial_d f^{LF}
\right\|_{L^2(V_b)}^2,
\end{equation}
where $V_b \subset \Omega$ denotes the $b$-th randomly sampled sub-volume in a training iteration, $B$ is the number of such sub-volumes, and $\partial_d$ denotes differentiation along spatial dimension $d$. In practice, spatial derivatives are approximated using first-order derivatives of a 3D Gaussian kernel
$G_{\sigma}(\mathbf{r}) = (2\pi\sigma^{2})^{-3/2} \exp\left(-\frac{\|\mathbf{r}\|^{2}}{2\sigma^{2}}\right)$,
with derivative filters
$k_{d}(\mathbf{r}) = -\frac{r_{d}}{\sigma^{2}} G_{\sigma}(\mathbf{r})$.
The discrete implementation becomes
\begin{equation}
\mathcal{L}_{grad} = \frac{1}{B} \sum_{b=1}^{B} \sum_{d \in \{x,y,z\}} 
\left\| 
\hat{V}_b * k_d - V_b * k_d 
\right\|_{F}^{2},
\end{equation}
where $\hat{V}_b$ and $V_b$ denote the predicted and ground-truth sub-volumes, respectively, and $\|\cdot\|_{F}$ denotes the Frobenius norm over the discrete voxel grid.

Gaussian derivatives provide noise-robust and rotationally symmetric gradient estimates suited to low-SNR MRI. Matching spatial derivatives enforces first-order consistency of the operator output, complementing the zero-order intensity constraint.

Once trained, the synthesized LF volume is obtained by evaluating the learned operator on the discrete sampling grid: $ \hat{I}^{LF}[i,j,k] = \hat{f}^{LF}(\mathbf{x}_{i,j,k}).$

\section{Experiments}
\subsection{Low Field MRI Synthesis}
A high-fidelity HF-to-LF transformation is essential for reliable in-silico retrospective clinical trials for novel LF device development, where only a small cohort of paired HF–LF scans—compared to the hundreds typically required in prospective studies~\cite{mazurek2021portable}—is available to learn the mapping. In this study, we rigorously evaluate the synthesis performance of H2LO against related methods.

\noindent \textbf{Dataset and Metrics.}
We train and evaluate on different partitions of the ds006557 dataset~\cite{vavsa2025ultra}, which consists of 23 subjects scanned at both high-field (3T GE) and low-field (64mT Hyperfine) conditions. Both T1w and T2w contrasts are available, with one T1w subject excluded due to missing data. All volumes are skull-stripped using SynthStrip~\cite{hoopes2022synthstrip} and registered to a common space via ANTs~\cite{tustison2021antsx}, yielding paired volumes of size $208 \times 256 \times 256$. Each volume is normalized to $[0, 1]$ via max normalization. We perform 5-fold cross-validation, with each fold using 13-14 subjects for training, 2 for validation, and 7 for testing. Reconstruction quality is measured using peak signal-to-noise ratio (PSNR, in  dB), structural similarity index (SSIM, in \%), and normalized cross-correlation (NCC) between  reconstructed and ground-truth images. Additionally, histogram-based metrics are used to assess intensity-distribution fidelity, including Wasserstein distance (WASS), histogram NCC (HNCC), Bhattacharyya distance (BHAT), and Jensen-Shannon (JS) divergence computed from intensity histograms. For histogram analysis, only foreground voxels are included using the ground-truth mask threshold ( $I_{HF} > 0.01 $).

\noindent \textbf{Comparison Methods.}
We compare our method with three categories of approaches.
(1) \textit{Image-to-image (I2I) translation}, which learns a direct mapping between HF and LF MRI voxels. We include Pix2Pix \cite{pix2pix2017} and DiffI2I \cite{xia2024diffi2i} as representative GAN-based and diffusion-based models, respectively.
(2) \textit{Physics-inspired LF MRI simulation} \cite{arnold2022simulated}, which approximates the LF acquisition process through noise modeling and spatial smoothing kernels.
(3) \textit{Conditional implicit neural representation (INR)}, which employs coordinate-based architectures similar to ours. Specifically, we adopt the 3D MRI implementation from \cite{wu2022arbitrary}, an extension of LIIF \cite{chen2021learning} designed for volumetric MRI data.

\noindent \textbf{Implementation.}
The branch network adopts a lightweight 3D CNN encoder~\cite{ledig2017photo} with five convolutional layers (kernel size $3^3$, channels 32--32--64--64--128) with ReLU activations, which produces a spatially-resolved feature map of dimension $P=128$. The trunk network is a 4-layer SIREN MLP (width 256, $\omega_0=30$) that maps normalized 3D coordinates to $P$ basis function values. For $\mathcal{L}_{1}$, $N$ is set to 8000. For $\mathcal{L}_{grad}$, we set $\sigma=1.0$ with a $5 \times 5 \times 5$ kernel based on hyperparameter search on validation sets and $B$ is set to 1. We train for 500 epochs using Adam with an initial learning rate of $10^{-4}$ and cosine annealing to $10^{-6}$. 
\vspace{-5pt}

\subsection{Low Field MRI Enhancement}
In this study, we examine the downstream utility of H2LO as a data engine for LF MRI enhancement, showcasing its ability to support LF algorithm development. We use an existing public HF dataset and simulate corresponding LF MRI using our trained HLFO. Generated HF-LF pairs thereby augment supervised LF MRI enhancement models. 

\noindent\textbf{Datasets and Metrics.}
The HF dataset is preprocessed 3T images from Open-neuro ds005752~\cite{ds005752:2.1.0}, which contains 184 subjects with both T1w and T2w scans.  These were split into 147/29 subjects for train/validation using a fixed random seed and an 80/20 partition. For each subject, LF counterparts for each HF structural MR volumes were generated (T1w and T2w available), with a typical voxel grid size of 208 $\times$ 256 $\times$ 256. All images are intensity-normalized to [0,1] and skull stripping is applied with SynthStrip. Real LF-HF pairs in ds006557 were retained for finetuning (using the training partition mentioned in Section 3.1) and evaluation (using the testing partition).

\noindent\textbf{Comparison Methods.}
(1) \textit{Real LF and Conventional Data Augmentations}: To increase data variation while preserving intensity consistency, we apply random paired in-plane augmentations (axial H,W flip and 90° rotations) to LF-HF training patches and keep the depth axis unchanged. (2) \textit{Synthetic (Pre)training and Real LF Finetuning}: We generate synthetic LF inputs from pretrained HF-to-LF models including ours and comparison methods listed in Section 3.1.

\noindent\textbf{Implementations.}
The PF-SR architecture~\cite{man2023deep} was adopted as the backbone network.
Since the present task focuses on same-resolution enhancement rather than super-resolution, the 3D sub-pixel convolution upsampling module was removed.
The final output is obtained via voxel-wise addition of the predicted residual and the original LF input. Pretraining was conducted on synthetic LF--HF image pairs with matched contrast, following the training protocol described in~\cite{man2023deep}. The model was subsequently fine-tuned on real LF--HF pairs using the same backbone and optimizer configuration. During fine-tuning, a cosine decay learning rate schedule was employed, with an initial learning rate of $2 \times 10^{-6}$, a 4-epoch warm-up phase starting from $0.5\times$ the base learning rate, and a minimum learning rate of $5 \times 10^{-7}$. Experiments are conducted on NVIDIA H200, A100 and RTX A5000 GPUs.

\section{Results}
\subsection{Low Field MRI Synthesis}
\begin{figure}[t]
    \centering
    \includegraphics[width=0.98\linewidth]{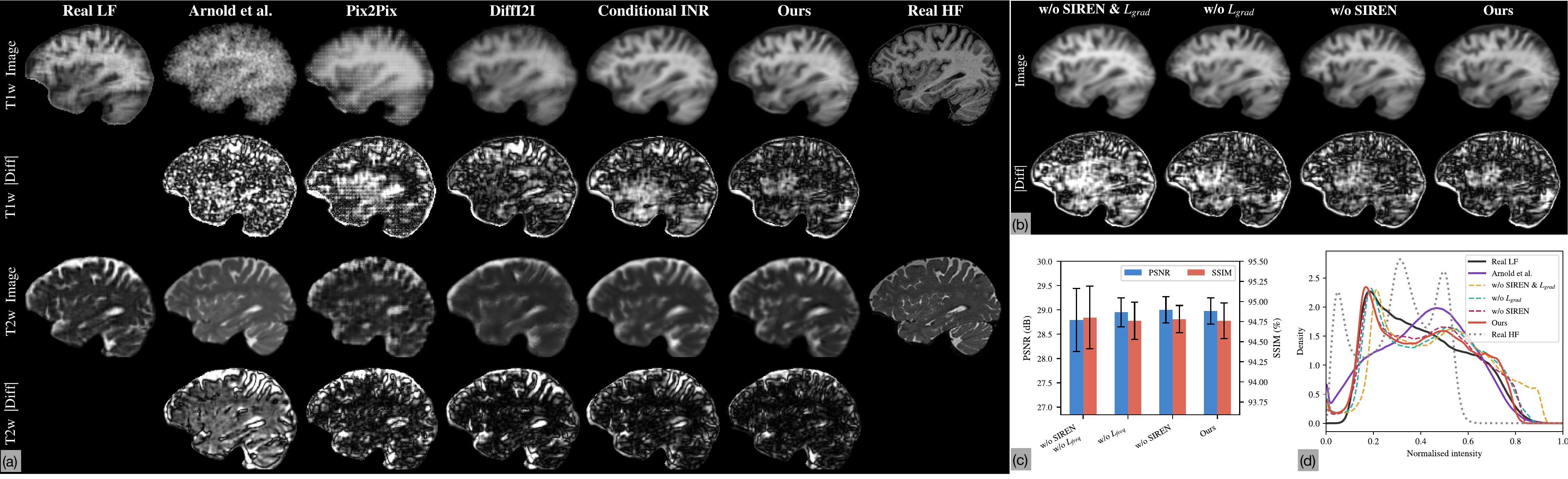}
    \caption{Visualization of generated LF MRIs and ablation studies. (a) Generated T1w (upper rows) and T2w (lower rows) LF images using multiple methods. Even rows visualize the absolute differences between simulated and real LF images. (b) Ablation of multiple components in our model. (c) Testing results of the 5-fold ablation study. (d) Comparison of histograms for LF images generated by Arnold et al. \cite{arnold2022simulated} and our method.}
    \label{fig:fig2}
\end{figure}

We present our main results of synthesis performance in Tab.~\ref{tab:syn} and Fig.~\ref{fig:fig2}(a) and (d). From Tab.~\ref{tab:syn}, our method consistently achieves the best performance across almost all metrics for both T1 and T2 contrasts. For T2 synthesis, Pix2Pix achieves the best results in part of the histogram-related statistics. However, the Pix2Pix column in Fig. \ref{fig:fig2}(a) shows that its visual results exhibit unnatural grid-like artifacts (which also occur in DiffI2I), and such artifacts are not desirable for high-fidelity retrospective studies on novel devices. From the same figure, we can also observe that our method achieves the highest visual alignment with real LF images. From Fig. \ref{fig:fig2}(a), Conditional INR produces overly smoothed LF images, while Arnold et al. simulate noise patterns but fail to properly model contrast changes, as reflected in the histogram comparison in Fig. \ref{fig:fig2}(d). In contrast, our method maintains both strong quantitative performance and more visually coherent and anatomically faithful results.

\begin{table*}[t]
\centering
\resizebox{0.98\textwidth}{!}{
\begin{tabular}{@{}l|ccccccc|r@{}}
\toprule
\multicolumn{9}{c}{\textbf{T1}} \\
\midrule
Method & PSNR(dB)$\uparrow$ & SSIM(\%)$\uparrow$ & NCC$\uparrow$ & WASS$\downarrow$ & HNCC$\uparrow$ & BHAT$\downarrow$ & JS$\downarrow$ & Param. \\
\midrule
HF Image & 23.52$\pm$0.44 & 89.96$\pm$0.34 & 0.9222$\pm$0.0035 & 0.1211$\pm$0.0107 & 0.5130$\pm$0.0436 & 0.2148$\pm$0.0218 & 0.1495$\pm$0.0130 & -- \\
Arnold et. al. \cite{arnold2022simulated} & 14.97$\pm$1.16 & 87.80$\pm$1.56 & 0.9466$\pm$0.0069 & 0.1099$\pm$0.0103 & 0.5759$\pm$0.0423 & 0.1009$\pm$0.0231 & 0.0823$\pm$0.0160 & 4 \\
Pix2Pix \cite{pix2pix2017} & 26.46$\pm$0.33 & 92.25$\pm$0.27 & 0.9561$\pm$0.0032 & 0.0403$\pm$0.0046 & 0.8302$\pm$0.0530 & 0.0448$\pm$0.0106 & 0.0358$\pm$0.0080 & 4.86M \\
DiffI2I \cite{xia2024diffi2i} & 27.16$\pm$0.54 & 92.50$\pm$0.83 & 0.9609$\pm$0.0054 & 0.0330$\pm$0.0046 & \underline{0.9028$\pm$0.0145} & 0.0364$\pm$0.0065 & 0.0287$\pm$0.0047 & 16.39M \\
Conditional INR \cite{wu2022arbitrary} & \underline{27.59$\pm$0.33} & \underline{93.90$\pm$0.32} & \underline{0.9650$\pm$0.0029} & \underline{0.0327$\pm$0.0031} & 0.8642$\pm$0.0132 & \underline{0.0426$\pm$0.0055} & \underline{0.0342$\pm$0.0043} & 0.85M \\
Ours & \textbf{28.98$\pm$0.27} & \textbf{94.76$\pm$0.22} & \textbf{0.9749$\pm$0.0017} & \textbf{0.0288$\pm$0.0049} & \textbf{0.9294$\pm$0.0141} & \textbf{0.0218$\pm$0.0068} & \textbf{0.0186$\pm$0.0057} & 0.58M \\
\midrule
\multicolumn{9}{c}{\textbf{T2}} \\
\midrule
Method & PSNR(dB)$\uparrow$ & SSIM(\%)$\uparrow$ & NCC$\uparrow$ & WASS$\downarrow$ & HNCC$\uparrow$ & BHAT$\downarrow$ & JS$\downarrow$ & Param. \\
\midrule
HF Image & 24.36$\pm$0.28 & 90.14$\pm$0.35 & 0.8059$\pm$0.0090 & 0.0583$\pm$0.0017 & 0.4818$\pm$0.0479 & 0.1759$\pm$0.0103 & 0.1442$\pm$0.0081 & -- \\
Arnold et. al. \cite{arnold2022simulated} & 23.43$\pm$0.15 & 90.42$\pm$0.44 & 0.8297$\pm$0.0061 & 0.1094$\pm$0.0030 & 0.0415$\pm$0.0269 & 0.3626$\pm$0.0108 & 0.2709$\pm$0.0067 & 4 \\
Pix2Pix \cite{pix2pix2017} & 25.45$\pm$0.28 & 91.42$\pm$0.46 & 0.8611$\pm$0.0095 & \underline{0.0259$\pm$0.0031} & \textbf{0.9609$\pm$0.0124} & \textbf{0.0212$\pm$0.0043} & \textbf{0.0180$\pm$0.0036} & 4.86M \\
DiffI2I \cite{xia2024diffi2i} & 26.88$\pm$0.23 & 92.33$\pm$0.36 & 0.8937$\pm$0.0074 & 0.0364$\pm$0.0037 & 0.9017$\pm$0.0203 & 0.0486$\pm$0.0056 & 0.0430$\pm$0.0049 & 16.39M \\
Conditional INR \cite{wu2022arbitrary} & \underline{27.80$\pm$0.37} & \underline{93.78$\pm$0.43} & \underline{0.9147$\pm$0.0065} & 0.0260$\pm$0.0022 & 0.8729$\pm$0.0491 & 0.0504$\pm$0.0145 & 0.0439$\pm$0.0119 & 0.85M \\
Ours & \textbf{28.37$\pm$0.38} & \textbf{94.29$\pm$0.48} & \textbf{0.9257$\pm$0.0064} & \textbf{0.0244$\pm$0.0050} & \underline{0.9458$\pm$0.0094} & \underline{0.0221$\pm$0.0027} & \underline{0.0205$\pm$0.0024} & 0.58M \\
\bottomrule
\end{tabular}
}
\caption{Quantitative results of LF MRI synthesis. The mean ± standard deviation across five-fold cross-validation are reported. The \textbf{best} and \underline{second-best} methods are highlighted in bold and underline, respectively.}
\label{tab:syn}
\vspace{-9pt}
\end{table*}

\subsection{Low Field MRI Enhancement}

Table~\ref{tab:enhance} summarizes the enhancement performance under different data augmentation strategies. Intensity-preserving conventional augmentation (flip \& rotation) does not improve performance over real-only training, indicating that simple transformations are insufficient to compensate for limited LF data. In contrast, incorporating synthetic LF images consistently maintains or improves performance. Comparison across synthesis methods suggests that effective LF modeling requires both contrast and noise degradation modeling. Arnold et al. primarily simulate noise, providing texture cues but lacking accurate contrast modeling. Conditional INR captures contrast shifts but produces overly smooth results with limited noise characteristics. By modeling both contrast changes and noise degradation, our method achieves the best result for both T1w and T2w MRI.

\begin{table}[t]
  \centering
  \resizebox{0.6\linewidth}{!}{
  \begin{tabular}{@{}l|ll|ll|l|l@{}}
  \toprule
                         & \multicolumn{2}{c|}{T1}       &
    \multicolumn{2}{c|}{T2}       & &      \\ \midrule
  Method   & PSNR(dB) & SSIM(\%) & PSNR(dB) & SSIM(\%)  & Subjects  & Epoch  \\
  \midrule
  Real LF             &30.12 &  92.93 & 29.26 & 92.61 & 13 & 200 \\
  Flip \& Rotate      & 29.49 & 92.06 & 28.50 & 91.98 & 13 (*8) & 400 \\
  Arnold et. al. \cite{arnold2022simulated} & 30.37  & 93.48 & 29.39 & 92.75 & 13+147 & 150+50\\
  Pix2Pix \cite{pix2pix2017} & \underline{30.39} & \underline{93.77} & \underline{29.59} & \underline{93.17} & 13+147 & 150+50\\
  DiffI2I \cite{xia2024diffi2i}& 30.27 & 93.33 & 29.52 & 93.01 & 13+147 & 150+50 \\
  Conditional INR \cite{wu2022arbitrary}  &  30.10 & 92.97 & 29.34 & 92.97 & 13+147 &150+50 \\
  Ours         &  \textbf{30.55}  & \textbf{93.71} & \textbf{30.02} & \textbf{93.72} & 13+147 & 150+50\\ \bottomrule
  \end{tabular}}
\caption{Quantitative result of LF MRI enhancement augmented with synthetic data. The \textbf{best} and \underline{second-best} methods are highlighted in bold and underline, respectively.}
\label{tab:enhance}
\vspace{-10pt}
\end{table}
  
\subsection{Ablation Study}
The ablation results for T1 synthesis are shown in Fig.~\ref{fig:fig2}(b)–(d). Visual comparisons in Fig.~\ref{fig:fig2}(b) indicate that the high-frequency gradient loss ($\mathcal{L}_{freq}$) improves local detail reconstruction and yields sharper LF textures. Histogram analysis in Fig.~\ref{fig:fig2}(b) and (d) further shows that incorporating SIREN enhances contrast modeling and contributes to finer structural representation. Quantitative results are reported in Fig.~\ref{fig:fig2}(c). Interestingly, ours w/o. SIREN achieves the highest PSNR and SSIM, while the full model ranks second. However, visual inspection shown in Fig.~\ref{fig:fig2}(b) suggests that the full model better preserves subtle anatomical details and produces more realistic LF characteristics.

\section{Conclusion and Future Work}
We present H2LO, a neural operator–based framework for subject-specific low-field MRI synthesis from high-field acquisitions. By formulating HF-to-LF translation as an operator learning problem, H2LO captures contrast shifts and high-frequency texture changes induced by field strength variation. Experiments on T1w and T2w MRI demonstrate improved synthesis fidelity over conventional simulators and representative image-to-image translation models. Moreover, synthetic LF images generated by H2LO enhance downstream LF image enhancement performance, supporting its use for virtual clinical evaluation and algorithm development in emerging low-field MRI systems.

A limitation of this study is that it includes only healthy subjects. While this enables controlled evaluation, generalization to pathological cases remains to be established, as lesions or atrophy may alter the HF-to-LF mapping. Future work will extend the framework to clinical cohorts with diverse pathologies. In addition, the study is conducted on a limited number of paired subjects, reflecting current data availability and aligning with virtual clinical trial settings. To ensure methodological robustness, we mitigate the limited sample size through 5-fold cross-validation. Future work will validate the proposed method on larger and more diverse cohorts as such data become available.

\bibliographystyle{plain}
\bibliography{ref}
\end{document}